\documentclass[letterpaper, 10 pt, conference]{ieeeconf}  
\IEEEoverridecommandlockouts                             
\usepackage{lineno} 
\usepackage{bm}

\usepackage{amssymb}
\usepackage[ruled,vlined]{algorithm2e}
\usepackage{cite}
\usepackage{graphicx}
\usepackage{amsmath}
\usepackage{array}
\usepackage{stfloats}
\usepackage{url}
\usepackage{xcolor}
\usepackage{comment}
\usepackage[hidelinks]{hyperref}

\title{\LARGE \bf
Cooperative Energy Management of HVAC via Transactive Energy
}

\author{Qing~Yang and Hao~Wang*
\thanks{Qing Yang is with the Blockchain Technology Research Center (BTRC) and the College of Electronics and Information Engineering (CEI), Shenzhen University, Shenzhen, Guangdong Province, China
        {\tt\small yang.qing@szu.edu.cn}}%
\thanks{Hao Wang is with the Department of Data Science and Artificial Intelligence, Faculty of Information Technology, Monash University, Melbourne, VIC 3800, Australia
        {\tt\small hao.wang2@monash.edu}}%
\thanks{This work is in part supported by the National Natural Science Foundation of China (project 61901280) and the FIT Academic Staff Funding of Monash University.}
\thanks{*Corresponding author: Hao Wang.}
}
\begin{document}
\maketitle
\thispagestyle{empty}
\pagestyle{empty}

\begin{abstract}
Heating, Ventilation, and Air Conditioning (HVAC) energy consumption accounts for a significant part of the total energy consumption of buildings and households. The ubiquitous adoption of distributed renewable energy and smart meters helps to decarbonize the HVAC energy consumption and improve energy efficiency. However, how to scale up HVAC energy management for a group of users while persevering users' privacy remains a big challenge. In this work, we utilize the concept of transactive energy to build a cooperative energy management system for independent HVAC units in a distributed manner. Specifically, we develop a distributed energy trading algorithm that consists of two layers based on the alternating direction method of multipliers method. The distributed energy trading algorithm achieves optimal trading performance and also preserves users' privacy. Furthermore, we evaluate the performance of the distributed trading algorithm by extensive simulations with real-world data. Simulation results show that the energy trading algorithm converges fast and the cooperative energy platform reduces the user's individual cost by up to 50\% and lowers the overall cost of all users by 23\%.

\end{abstract}

\section{INTRODUCTION}\label{sec:intro}

Heating ventilation and air conditioning (HVAC) is an indispensable component of modern buildings; however, HVAC energy use accounts for a significant portion of energy consumption in buildings and households \cite{none2016annual}. Recent research interests have been focusing on how to decarbonize the HVAC energy consumption. The ubiquitous deployment of smart meters and renewable energy generators enables us to develop energy management systems for HVAC units powered by a high penetration of renewable energy. As an essential part of the smart city, residential users (e.g., smart homes) \cite{tushar2014three} are expected to manage their energy consumption to improve the energy efficiency of appliances (and the HVAC in particular). This paper aims to study the following key problem: can residential users further reduce their HVAC expenses by merely utilizing the existing grid and renewable energy? To this end, transactive energy emerges as a prospective solution \cite{wang2016incentivizing} in both smart grid academia and industry.

In smart homes, the energy management system schedules the energy usage of appliances including HVAC to reduce energy costs yet keeps the indoor comfortable \cite{fukuta2015}. To further improve the energy efficiency, users are encouraged to cooperate in the energy management of their renewable supply, HVAC, and other load. By enabling users to share surplus energy and cooperatively schedule their energy usage, cooperative energy management can effectively reduce the individual user's cost and enhances the efficiency of the system. However, the following challenges must be addressed before a cooperative energy management system that involves multiple users in a community. First, the cooperative energy management system should improve the system efficiency while incentivizing the users to join the trading by providing them enough benefit. Second, privacy-preserving is critical for cooperative energy management since users' power usage information can be exploited by malicious parties for their own interests.

Recently, many research works on HVAC energy management have been published to address the above challenges. A data-driven modeling method for HVAC was proposed in \cite{kusiak2010modeling} to achieve the optimal HVAC energy scheduling. In \cite{fong2009system}, the authors suggested a robust evolutionary algorithm to optimize the power consumption of centralized HVAC. In \cite{nguyen2014energy}, the authors investigated the solar-assisted HVAC model and proposed a stochastic optimization method to reduce energy consumption. However, references \cite{kusiak2010modeling, fong2009system, nguyen2014energy} only studied standalone energy management of HVAC in a single house. A cooperative energy management that involves multiple households was studied in \cite{luna2016cooperative}, in which the authors employed a mixed-integer linear programming method but assumed a fully centralized model that causes privacy concerns. Reference \cite{pnnl} reports the vulnerability of users' privacy in transactive energy, and \cite{eibl2014influence} presented a method to exploit privacy information from users' smart meter data. The trade-off between users' privacy and the performance of HVAC energy management is modeled and analyzed in \cite{jia2017privacy}, and the authors suggested a noise injection method to enhance privacy.

In this paper, we designed a cooperative energy management system for HVAC with a novel distributed P2P (peer-to-peer) energy trading algorithm to address the challenges as mentioned earlier. The main contributions of this work are as follows.
\begin{enumerate}
  \item \textit{We develop an efficient and trustable cooperative energy management platform for HVAC based on transactive energy.} The developed platform  enables energy trading among users and handles the supply/demand matching while preserving users' privacy. 
  \item \textit{We design a P2P energy trading algorithm that improves the system efficiency and minimizes users' HVAC costs.} The proposed energy trading algorithm enables participated users to interact in a distributed manner to gain benefits from trading, and also does not require users to reveal private parameters.
  \item \textit{We evaluate the distributed P2P energy trading algorithm with extensive simulation using real-world data.} To test the feasibility of the energy trading algorithm in practice, we validate the trading platform and algorithm using realistic renewable and load data. The results show that the system cost is reduced by 23\% and the users' costs are reduced by up to 50\%.
\end{enumerate}

The rest of the paper is organized as follows. Section~\ref{sec:model} introduces the HVAC energy management model on the transactive energy platform. Section~\ref{sec:formulation} formulates an optimization problem for the cooperative energy system. Section~\ref{sec:solution} elaborates on the design of the distributed P2P energy trading algorithm for HVAC. Section~\ref{sec:eval} evaluates the proposed P2P energy trading system with extensive simulations and Section~\ref{sec:conclusion} concludes our work.

\section{SYSTEM MODEL}\label{sec:model}

\begin{figure}[!ht]
    \centering
    \includegraphics[width=8.7cm]{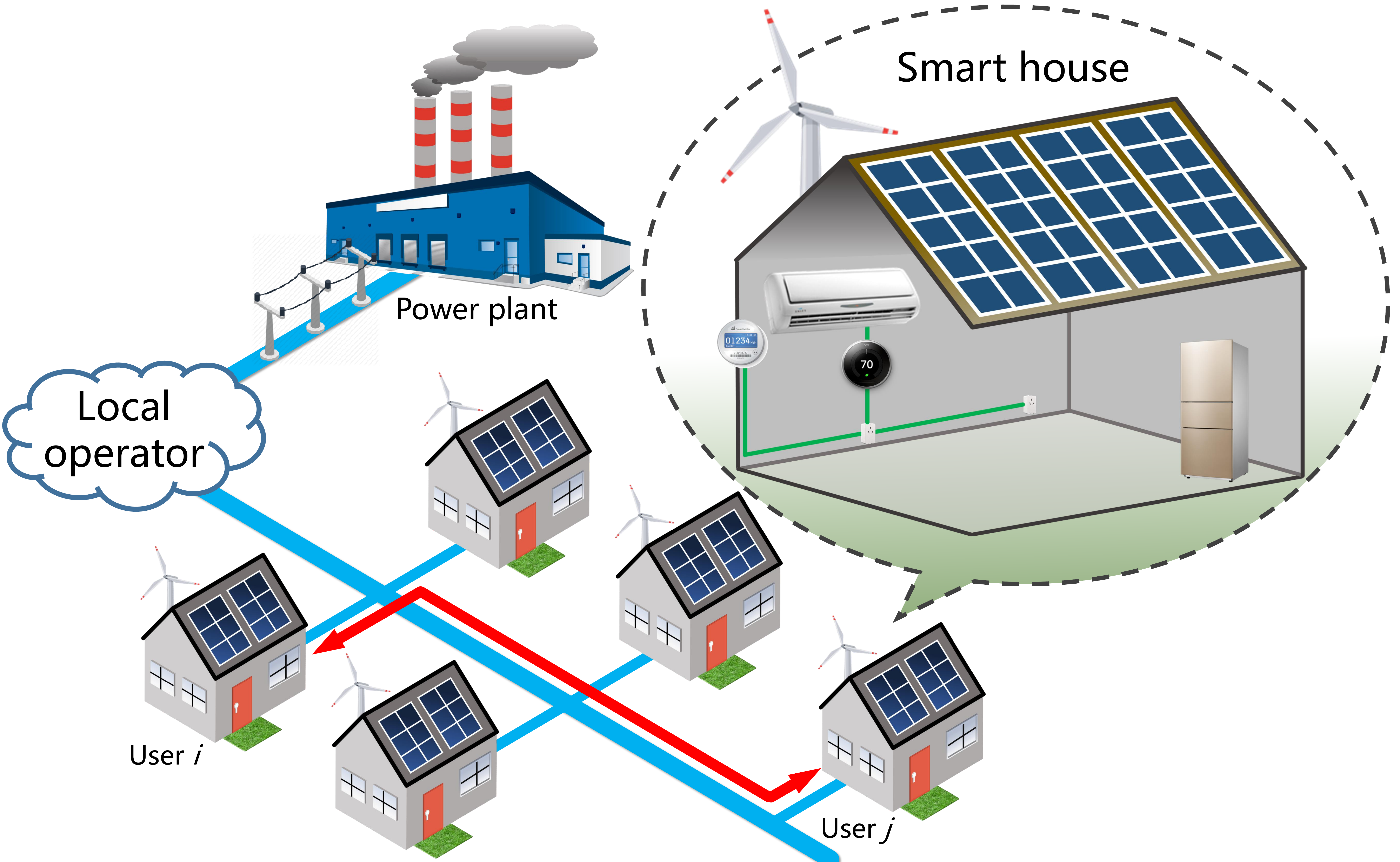}
    \caption{The system model of the transactive energy platform. The user's smart house is installed with renewable generation and appliances. The energy management system manages the HVAC according to the residential user's preference. The smart house can generate renewable energy, draw electricity from the grid, and trade energy with other smart houses. The energy management system is used to manage the HVAC energy consumption and run the P2P energy trading algorithm.}
    \label{f1:sysmod}
\end{figure}

We consider multiple houses interconnected via the power grid, as shown in Fig.~\ref{f1:sysmod}. We denote the users in this energy management model by the set $\mathcal{N} {=} \{1,..,N\}$. The system can be split into two layers: 1) the \textit{internal energy scheduling layer}, in which each user schedules the energy use for the HVAC; and 2) the \textit{external energy trading layer} facilities energy exchange among users to further utilize local renewables and reduce operational costs. Exchanging surplus energy with other users is encouraged during the operation of the energy management system. The residential users have diverse preferences for their household energy usage including HAVC (e.g., target temperature); therefore the users may benefit each other by trading energy in a peer-to-peer manner. The system model of the user's smart house is described in Fig.~\ref{f1:sysmod}.  

\subsection{Internal Energy Scheduling Layer}
Each residential user is equipped with renewable generation (e.g., solar panel and wind turbine), HVAC, and other appliances as shown in Fig.~\ref{f1:sysmod}. The model of the smart house has two parts. The first part is the  \textit{supply model} that includes the electricity from both the grid and local renewable energy; the second part is the \textit{load model} that consists of HVAC and other appliances. Let $\mathcal{H} {=} \{1,..., H \}$ denotes the operational horizon of the system. The load model of user $i$ describes the energy consumption of the user's HVAC and other appliances. Specifically, we denote the amount of energy that user $i$ purchases from the grid by  $p_i^{\text{G}}[t]$, and denote the renewable energy by $P_i^{\text{RE}}[t]$ in time slot $t$. Therefore, the energy supply terms $p_i^{\text{G}}[t]$ and $P_i^{\text{RE}}[t]$ should satisfy the following constraints:
    \begin{align}
            0 \leq p_i^{\text{RE}}[t] \leq P_i^{\text{RE}}[t], &~ \forall i \in \mathcal{N}, t \in \mathcal{H}, \label{constraint-load5} \\
            0 \leq  p_i^{\text{G}}[t] \leq P_i^{\text{G}}, &~ \forall i \in \mathcal{N}, t \in \mathcal{H}, \label{constraint-load6}
    \end{align}
where $p_i^{\text{RE}}[t]$ and $p_i^{\text{G}}[t]$ are non-negative and upper-bounded by the available renewable generation $P_i^{\text{RE}}[t]$ and power capacity $P_i^{\text{G}}$ from the grid per time slot, respectively. 

To model the realistic system, we adopt the two-part tariff billing for the electricity purchased from the grid. Hence, user $i$'s grid cost is
    \begin{align}
            C_i^{\text{G}} = \pi_1^{\text{G}} \sum \nolimits_{t\in \mathcal{H}} p_i^{\text{G}}[t] + \pi_2^{\text{G}} \max_{t\in \mathcal{H}}      p_i^{\text{G}}[t], \label{objective-supply} 
    \end{align}
where $\pi_1^{\text{G}}$ and $\pi_2^{\text{G}}$ denote the unit energy price and peak load price, respectively.
The grid cost consists of an energy charge $\pi_1^{\text{G}} \sum_{t\in \mathcal{H}} p_i^{\text{G}}[t]$ and a peak charge $\pi_2^{\text{G}} \max_{t\in \mathcal{H}} p_i^{\text{G}}[t]$. The peak charge is introduced to shave the peak load of the grid by penalizing the user's peak load.

The \textit{load model} includes the HVAC load and other inflexible loads such as the refrigerator and lighting. We let $P_i^{\text{IL}}[t]$ denote the inflexible load, which includes the power consumption of house appliances other than HVAC in time slot $t$. The HVAC system works to adjust the indoor temperature $T_i^{\text{IN}}[t]$ according to user $i$'s preference. In time slot $t$, the energy consumption of the heating or cooling process incurred by the HVAC system is denoted by $p_i^{\text{AC}}[t]$. We denote the outdoor temperature as $T_i^{\text{OUT}}[t]$, then the dynamics of the indoor temperature \cite{cui2019} is modeled as
    \begin{equation}
      \begin{aligned}
            T_i^{\text{IN}}[t] = T_i^{\text{IN}}[t-1] -& \frac{1}{C_i R_i} ( T_i^{\text{IN}}[t-1] - T_i^{\text{OUT}}[t] \\
            &+ \eta_i R_i p_i^{\text{AC}}[t] ),~ \forall i \in \mathcal{N}, t \in \mathcal{H}, \label{constraint-load1}
      \end{aligned}
    \end{equation}
where $C_i$ and $R_i$ denote the working parameters of the HVAC unit (and we let $C_i {=} 3.3$ and $R_i {=} 1.35$ in our simulations). The working mode of the HVAC is represented by the sign of $\eta_i$: $\eta_i$ is positive if the system is cooling, and negative if the system is heating. Through this work, we omit the installation cost and depreciation of all the appliances in the model since our focus is energy management.

Usually, user $i$ sets a preferred indoor temperature $T_i^{\text{REF}}$ for the HVAC system; thus any deviation from this value will discomfort the user. The discomfort of user $i$ can be modeled by the cost function below
    \begin{equation}
            C_i^{\text{AC}} = \beta_i^{\text{AC}} \sum \nolimits_{t \in \mathcal{H}} \left( T_i^{\text{IN}}[t] - T_i^{\text{REF}} \right)^{2}, ~ \forall i \in \mathcal{N}, \label{objective-load1}
    \end{equation}
where $\beta_i^{\text{AC}}$ denotes that sensitivity coefficient.

Note that the indoor temperature should also be within a range that human can tolerate, thus we have
    \begin{equation}
            \underline{T}_i^{\text{IN}} \leq  T_i^{\text{IN}}[t] \leq  \overline{T}_i^{\text{IN}}, ~ \forall i \in \mathcal{N}, t \in \mathcal{H}, \label{constraint-load2}
    \end{equation}
where the parameters $\underline{T}_i^{\text{IN}}$ and $\overline{T}_i^{\text{IN}}$ define the upper- and lower-bounds of user $i$'s tolerable temperature.

\subsection{External Energy Trading Layer}
Besides scheduling their internal energy, users can also exchange energy externally with each other via the cooperative energy management system. Because users have diverse HVAC use profiles and different renewable energy generations, they can exploit the diversities of supply and load profiles for mutual benefits by transactive energy.

In the cooperative energy management system shown in Fig.\ref{f1:sysmod}, user $i$ can trade with user $j \in \mathcal{N} \backslash i$ to exchange energy. The cooperative energy management platform provides a free and open marketplace for users to trade energy and allow participated users to earn profit. Let $p_{i,j}^{\text{ET}}[t]$ denotes the amount of electricity traded between user $i$ and user $j$ in time slot $t$. Note that a positive $p_{i,j}^{\text{ET}}[t]$ indicates that user $i$ purchases energy from user $j$, and negative $p_{i,j}^{\text{ET}}[t]$ indicates user $i$ sells energy to user $j$.

Without loss of generality, we omit the electricity loss during the power transfer over the grid, since the users are closely located to each other. To balance the load and supply over the grid, the following constraint applied for all the energy trading pairs:
    \begin{align}
        & p_{i,j}^{\text{ET}}[t] + p_{j,i}^{\text{ET}}[t] = 0,~\forall t \in \mathcal{H},~\forall i \in \mathcal{N},~\forall j \in \mathcal{N} \backslash i, \label{constraint-trading1}     \end{align}
which clears the energy-trading amount between each trading pair of users $i$ and $j$ in time slot $t$.

\section{PROBLEM FORMULATION}\label{sec:formulation}
In this section, we consider the HVAC energy management scenarios without and with energy trading among users. Specifically, in the non-cooperative scenario, all the users optimize their own energy scheduling independently without interacting with other users. In the cooperative scenario, users exchange energy with other users to purchase energy when in a deficit of local supply or sell extra energy to make profits. 

\subsection{Non-cooperative Scenario}
In the non-cooperative energy management scenario, the residential users schedule their energy usages individually. Specifically, users schedule their energy supply and energy consumption to minimize the operating costs. User $i$ uses the local renewable supply $p_i^{\text{RE}}[t]$ and also purchases from the grid $p_i^{\text{G}}[t]$ to serve the demand, consisting of HVAC load $p_i^{\text{AC}}[t]$ and the inflexible load $p_i^{\text{IL}}[t]$. To shorten the notation, we redefine the following variables as $\bm{p}_i^{\text{RE}} {=} \{ p_i^{\text{RE}}[t], \forall t {\in} \mathcal{H}\}$,
$\bm{p}_i^{\text{G}} {=} \{ p_i^{\text{G}}[t], \forall t {\in} \mathcal{H}\}$,
$\bm{p}_i^{\text{AC}} {=} \{ p_i^{\text{AC}}[t], \forall t {\in} \mathcal{H}\}$,
and
$\bm{p}_i^{\text{IL}} {=} \{ p_i^{\text{IL}}[t], \forall t {\in} \mathcal{H}\}$.

During the energy management process, user $i$'s power supply and demand must be equal in each time slot, which leads to the constraint
    \begin{align}
            p_i^{\text{RE}}[t] + p_i^{\text{G}}[t] = p_i^{\text{AC}}[t] + P_i^{\text{IL}}[t],
            ~ \forall i \in \mathcal{N}, t \in \mathcal{H}. \label{constraint-load11}
    \end{align}
The left-hand side of \eqref{constraint-load11} represents the total energy supply, and the right-hand side of \eqref{constraint-load11} denotes the total demand from the HVAC load $p_i^{\text{AC}}[t]$ and the inflexible load $P_i^{\text{IL}}[t]$.

For the non-cooperative energy management, the overall operating cost of user $i$ is
\begin{align}
             C_i^{\text{O}}(\bm{p}_i^{\text{G}}, \bm{p}_i^{\text{AC}}) 
             \triangleq
            C_i^{\text{G}}(\bm{p}_i^{\text{G}}) + C_i^{\text{AC}}(\bm{p}_i^{\text{AC}}), \label{objective-operatingcost}
\end{align}
where $C_i^{\text{G}}(\bm{p}_i^{\text{G}})$ denotes the electricity bill from the grid, and $C_i^{\text{AC}}(\bm{p}_i^{\text{AC}})$ denotes the the user's discomfort cost of HVAC.

The energy management system aims to minimize users' total cost in \eqref{objective-operatingcost}. Therefore, for a specific user $i$, the energy management problem can be formulated as 

\noindent\textbf{EMP$_i$}: User $i$'s Energy Management Problem.
    \begin{equation*}
        \begin{aligned}
            & \min
            && C_i^{\text{O}}(\bm{p}_i^{\text{G}}, \bm{p}_i^{\text{AC}}) \\
            & \text{subject to} 
            && \text{\eqref{constraint-load5}},\text{\eqref{constraint-load6}}, \text{\eqref{constraint-load1}},\text{\eqref{constraint-load2}},\text{\eqref{constraint-load11}}\\
            & \text{variables:} 
            && \left\{ \bm{p}_i^{\text{RE}}, \bm{p}_i^{\text{G}},  \bm{p}_i^{\text{AC}} \right\}.
        \end{aligned}
    \end{equation*}

Based on our analysis, user $i$ can locally solve the optimization problem in \textbf{EMP$_i$} since it is a standard convex optimization. We denote the optimal total cost in problem \textbf{EMP$_i$} by $\bar{C}_i^{\text{O}}$ and let this value serve as a benchmark cost for the comparison with the costs in the cooperative scenario.

\subsection{Cooperative Scenario}
Unlike the non-cooperative scenario, users not only schedule their internal energy supply and demand but also exchange energy externally with other users when needed in the cooperative scenario. Users with surplus renewable energy can sell their energy to other users via the grid; also, users can purchase energy from their neighbors other than the grid to lower the total costs. Taken the energy trading into account, the load balance constraint becomes
    \begin{equation}
        \begin{aligned}
            p_i^{\text{RE}}[t] {+} p_i^{\text{G}}[t] {+} & \sum_{j \in \mathcal{N} \backslash i} p_{i,j}^{\text{ET}}[t] {=} p_i^{\text{AC}}[t] {+} P_i^{\text{IL}}[t],  \forall i {\in} \mathcal{N}, t {\in} \mathcal{H}, \label{constraint-load12}
        \end{aligned}
    \end{equation}
where the summation $\sum_{j \in \mathcal{N} \backslash i} p_{i,j}^{\text{ET}}[t]$ is the total energy traded between user $i$ and all other users $j \in \mathcal{N} \backslash i$ in time slot $t$. 

We assume that the operator of the energy management system does not charge any fees to the users. Therefore, users who purchase energy will pay their counterparts at a time-dependent unit price $\pi[t]$ set by the trading platform. Therefore, user $i$'s cost during the energy trading process is
    \begin{align}
        C_{i}^{\text{ET}} (\boldsymbol{p}_i^{\text{ET}}) = \sum \nolimits_{t \in \mathcal{H}} \left( \pi[t] \sum \nolimits_{j \in \mathcal{N} \backslash i} p_{i,j}^{\text{ET}}[t] \right), \label{objective-tradingpayment}
    \end{align}
where $\boldsymbol{p}_i^{\text{ET}}$ is the short notation for $\{ p_{i,j}^{\text{ET}}[t],~\forall t \in \mathcal{H},~\forall j \in \mathcal{N} \backslash i \}$.

In the cooperative energy management system, the system overall cost consists of user's individual operating cost $C_i^{\text{O}}(\bm{p}_i^{\text{G}}, \bm{p}_i^{\text{AC}})$ and the trading cost $C_{i}^{\text{ET}} (\boldsymbol{p}_i^{\text{ET}})$. Therefore, we formulate the cooperative energy management problem \textbf{CEMP} for the platform as follows.

\noindent\textbf{CEMP}: Cooperative Energy Management Problem
    \begin{equation*}
        \begin{aligned}
            & \min && \sum_{i\in\mathcal{N}} 
            \left[ 
            C_i^{\text{O}}(\bm{p}_i^{\text{G}}, \bm{p}_i^{\text{AC}})
            + C_{i}^{\text{ET}} (\boldsymbol{p}_i^{\text{ET}}) 
            \right] \\
            & \text{subject to} 
            && \text{\eqref{constraint-load5}}, \text{\eqref{constraint-load6}},
            ~\text{\eqref{constraint-load1}},
            ~\text{\eqref{constraint-load2}},
            ~\text{\eqref{constraint-trading1}},
            ~\text{\eqref{constraint-load12}}\\
            & \text{variables:}
            &&\{ \bm{p}_i^{\text{RE}}, \bm{p}_i^{\text{G}}, \bm{p}_i^{\text{AC}}, \bm{p}_i^{\text{ET}}, ~i \in \mathcal{N} \}.
        \end{aligned} 
    \end{equation*}
We see that Problem \textbf{CEMP} jointly solves the optimal internal energy scheduling (of renewable and HVAC load) and the optimal external energy trading for all the users. However, solving \textbf{CEMP} in a centralized manner may cause severe privacy concerns because the users have to reveal all their parameters to the platform. To preserve users' privacy and enable cooperative energy management, we aim to solve Problem \textbf{CEMP} in a distributed manner in Section \ref{sec:solution}. 

\section{ALGORITHM DESIGN}\label{sec:solution}
Distributed solutions have been developed in microgrid energy management \cite{wang2016incentivizing}, data center demand response \cite{wang2016renewable}, and electric vehicles \cite{rivera2016distributed}. In our work, though users are willing to trade energy for benefits, they have the concern of sharing their private data and system parameters with others. Motivated by the need for privacy-preserving, we design a distributed trading algorithm to solve Problem \textbf{CEMP}. In the distributed algorithm, the users optimize their internal energy scheduling of renewable and HVAC load and communicate with others to determine the external energy trading. Therefore, users do not share their private data in the energy-trading process.

First, we employ the alternating direction method of multipliers (ADMM) method \cite{boyd2011distributed} to solve Problem \textbf{CEMP}, as ADMM has a good convergence for convex optimization problems with non-strictly convex objective functions. Here, we introduce the auxiliary variables $\hat{\boldsymbol{p}}_i^{\text{ET}} {=} \{\hat{p}_{i,j}^{\text{ET}}[t], \forall j \in \mathcal{N} \backslash i, \forall t \in \mathcal{H}\}$ for energy trading decisions of user $i$, and thus rewrite constraint \eqref{constraint-trading1} to its equivalent constraints
    \begin{align}
        & \hat{p}_{i,j}^{\text{ET}}[t] = p_{i,j}^{\text{ET}}[t],~\forall j \in \mathcal{N} \backslash i,~\forall i \in \mathcal{N},~\forall t \in \mathcal{H}, \label{constraint-auxiliary1}\\
        & \hat{p}_{i,j}^{\text{ET}}[t] + \hat{p}_{j,i}^{\text{ET}}[t] = 0,~\forall j \in \mathcal{N} \backslash i,~\forall i \in \mathcal{N},~\forall t \in \mathcal{H}. \label{constraint-auxiliary2}
    \end{align}

To solve Problem \textbf{CEMP}, we decompose it into a lower-level problem and an higher-level problem. To facilitate the decomposition, we introduce dual variables $\boldsymbol{\lambda} = \{ \bm{\lambda}_{i},~\forall i \in \mathcal{N} \}$ where $\bm{\lambda}_{i} = \{ \lambda_{i,j}[t],~\forall j \in \mathcal{N} \backslash i,~t \in \mathcal{H}\}$ for constraints \eqref{constraint-auxiliary1}. We then obtain the augmented Lagrangian for Problem \textbf{CEMP} as
    \begin{equation*}
    \begin{aligned}
           L = \sum_{i\in\mathcal{N}} 
            \big[ C_i^{\text{O}}(\bm{p}_i^{\text{G}}, & \bm{p}_i^{\text{AC}}) + C_{i}^{\text{ET}} (\bm{p}_i^{\text{ET}}) \big] \\
        + \sum_{i\in\mathcal{N}} & \sum_{j \in \mathcal{N} \backslash i} \sum_{t\in\mathcal{H}} 
        \Big[ \frac{\rho}{2} \left( \hat{p}_{i,j}^{\text{ET}}[t] - p_{i,j}^{\text{ET}}[t] \right)^{2} \\
        & + \lambda_{i,j}[t] ( \hat{p}_{i,j}^{\text{ET}}[t] - p_{i,j}^{\text{ET}}[t]) \Big], 
    \end{aligned}    
    \end{equation*}
where $\rho >0$ is a coefficient for the quadratic penalty of constraint \eqref{constraint-auxiliary1}.
    
Next, we split the optimization of \textbf{CEMP} into a lower-level problem and an higher-level problem. In the lower-level problem, users minimize their own operating costs in parallel when the dual variables $\boldsymbol{\lambda}$ and auxiliary variables $\{ \boldsymbol{\hat{p}}_i^{\text{ET}}, \forall i {\in} \mathcal{N} \}$ are given. The higher-level problem updates the auxiliary variables and dual variables based on the trading decisions submitted by users $\{ \boldsymbol{p}_i^{\text{ET}}, \forall i {\in} \mathcal{N} \}$. The trading algorithm iterates between the lower-level problem and the higher-level problem until it converges to the optimal solution. 

Specifically, in the lower-level problem, given the dual variables $\lambda_{i,j}[t]$ and auxiliary variables $\{ \hat{p}_{i,j}^{\text{ET}}[t], j {\in} \mathcal{N} \backslash i \}$, user $i$ solves the following optimization problem:
    
\textbf{LLP1$_i$}: Lower-level problem of \textbf{CEMP}.
    \begin{equation*}
        \begin{aligned}
            & \min 
            &&     C_i^{\text{O}}(\bm{p}_i^{\text{G}}, \bm{p}_i^{\text{AC}})
            + C_{i}^{\text{ET}} (\boldsymbol{p}_i^{\text{ET}}) \\
            &&& {+} \sum_{j \in \mathcal{N} \backslash i} \sum_{t\in\mathcal{H}} \left[ \frac{\rho}{2} \left( \hat{p}_{i,j}^{\text{ET}}[t] {-} p_{i,j}^{\text{ET}}[t] \right)^{2} - \lambda_{i,j}[t] p_{i,j}^{\text{ET}}[t] \right] \\
            & \text{subject to}
            && \text{\eqref{constraint-load5}}-\text{\eqref{constraint-load6}},
            ~\text{\eqref{constraint-load1}},
            ~\text{\eqref{constraint-load2}},
            ~\text{\eqref{constraint-load12}} \\
            & \text{variables:}
            && \{ \bm{p}_i^{\text{RE}}, \bm{p}_i^{\text{G}}, \bm{p}_i^{\text{AC}}, \bm{p}_i^{\text{ET}} \}.
        \end{aligned} 
    \end{equation*}
By solving \textbf{LLP1}, user $i$ obtains the energy trading decision $\bm{p}_i^{\text{ET}}$ and sends to the higher-level problem for the iteration.
    
The higher-level problem solves the dual variables $\boldsymbol{\lambda}$ and auxiliary variables $\boldsymbol{\hat{p}}_i^{\text{ET}}$ for all users $i \in \mathcal{N}$. The higher-level problem of \textbf{CEMP} is formulated as follows:
    
\textbf{HLP1}: higher-level problem of \textbf{CEMP}
    \begin{equation*}
        \begin{aligned}
            & \min  && \sum_{i\in\mathcal{N}} \sum_{j \in \mathcal{N} \backslash i} \sum_{t\in\mathcal{H}} 
        \left[ \frac{\rho}{2} \left( \hat{p}_{i,j}^{\text{ET}}[t] {-} p_{i,j}^{\text{ET}}[t] \right)^{2} 
        {+} \lambda_{i,j}[t] \hat{p}_{i,j}^{\text{ET}}[t] \right] \\
            & \text{subject to} 
            && \text{\eqref{constraint-auxiliary2}} \\
            & \text{variables:} 
            && \{ \hat{\bm{p}}_i^{\text{ET}},~i \in \mathcal{N} \},
        \end{aligned} 
    \end{equation*}
and solving \textbf{HLP1} yields the optimal auxiliary variables
        \begin{align}
            \begin{split}
                & \hat{p}_{i,j}^{\text{ET}}[t] {=} {-}\hat{p}_{j,i}^{\text{ET}}[t]
                =  \frac{\rho \left( p_{i,j}^{\text{ET}}[t] {-} p_{j,i}^{\text{ET}}[t] \right) - \left( \lambda_{i,j}[t] - \lambda_{j,i}[t] \right) }{2 \rho}, \label{updateenergy}
            \end{split}
        \end{align}
and the update rule for the dual variables as
    \begin{align}
        \lambda_{i,j}[t+1] = \lambda_{i,j}[t] + \rho \left( \hat{p}_{i,j}^{\text{ET}}[t] - p_{i,j}^{\text{ET}}[t] \right). \label{updatelambda}
    \end{align}

\begin{algorithm}[ht]
     \caption{Distributed transactive energy trading}
     \label{alg1} 
     \SetAlgoLined
     \textbf{Initialization}: Index $k {\leftarrow} 1$, tolerance $\epsilon {\leftarrow} 0.1$, stepsize $\rho(0) {\leftarrow} 1$, multipliers $\boldsymbol{\lambda} {\leftarrow} \boldsymbol{0}$;
    
    \While{$\sum_{i \in \mathcal{N}} \parallel \bm{\hat{p}}_i^{\mathrm{ET}}(k) - \bm{p}_i^{\mathrm{ET}}(k) \parallel > \epsilon $}{
    \hspace*{-1.1em}$\neg$ \For{$i \in \mathcal{N}$}{
        \hspace*{-1.1em}$\neg$ User $i$ solves Problem \textbf{LLP1$_i$} based on $\hat{\bm{p}}_i^{\text{ET}}(k-1)$ and $\bm{\lambda}_i(k-1)$;
        
        \hspace*{-1.1em}$\neg$ User $i$ sends the energy-trading decisions $\bm{p}_i^{\text{ET}}(k)$ to Problem \textbf{HLP1};
    }
    
    \hspace*{-1.1em}$\neg$ \textbf{HLP1} updates $\bm{\hat{p}}_{i}^{\text{ET}}(k),\forall i \in \mathcal{N}$ and $\bm{\lambda}(k)$ according to (\ref{updateenergy}) and (\ref{updatelambda});
    
    \hspace*{-1.1em}$\neg$ Update the iteration index $k \leftarrow k+1$;
    }
 \textbf{Results}:Optimal HAVC management and energy trading schedule, i.e.,  $\bm{p}_i^{\text{RE},\ast}$, $\bm{p}_i^{\text{G},\ast}$, $\bm{p}_i^{\text{AC},\ast}$, $\bm{p}_i^{\text{ET},\ast}$.
\end{algorithm}

We present the distributed transactive energy algorithm in Algorithm~\ref{alg1}. Specifically, the algorithm solves the higher-level and lower-level problems for Problem \textbf{CEMP} in an iterative fashion. Each user determines the internal energy scheduling and external energy trading in Problem \textbf{LLP1$_i$} without sharing any private data. Instead, each user only needs to report its trading demand to the higher-level problem. The higher-level problem receives all the users' trading decisions to update the auxiliary and dual variables in \eqref{updateenergy} and \eqref{updatelambda}. Then the higher-level problem broadcasts the updated information to all the users. According to \cite{boyd2011distributed}, our developed distributed algorithm (i.e., representing the higher-level and lower-level problems) converges to the optimal solution of Problem \textbf{CEMP} when we choose the stepsize $\rho(k) = 1/k$, in which $k$ denotes the number of iteration.

\begin{figure*}[!b]
    \centering
    \includegraphics[width=12cm]{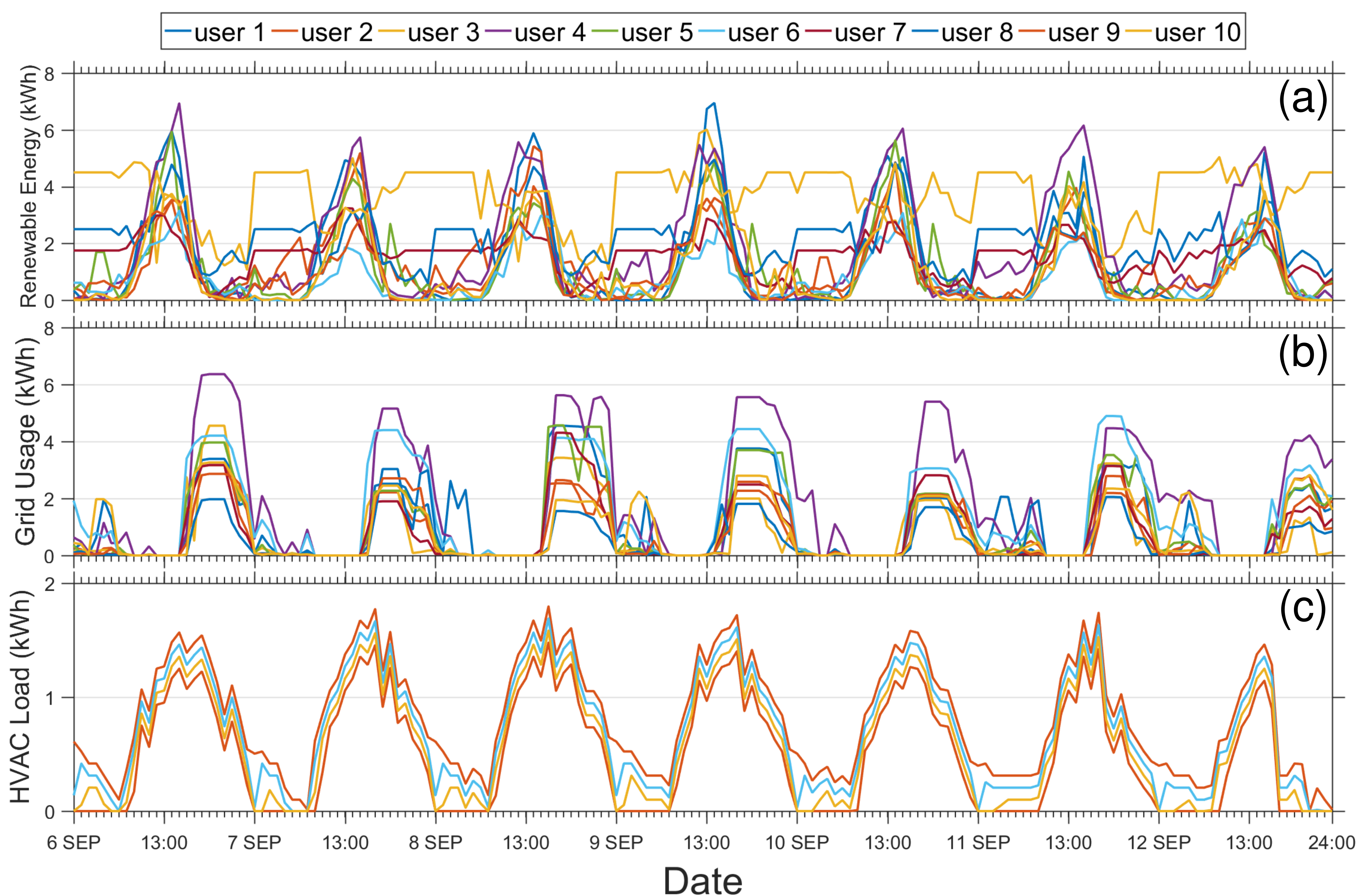}
    \caption{The optimal energy scheduling of $10$ users for (a) renewable energy supply, (b) energy drawn from the grid, (c) energy consumed by HVAC over one week. }
    \label{f:usage}
\end{figure*}

\section{SIMULATION RESULTS}\label{sec:eval}

In this section, we evaluate the cooperative HAVC energy management platform by extensive numerical simulations with real data of renewable generation \cite{wang2016microgrid} and load \cite{pecan}. The simulation data include users' daily electricity usage, renewable energy generation (e.g., rooftop solar panels and wind turbines), and outdoor temperature during September 6-12, 2016. To validate the energy trading algorithm developed in Section~\ref{sec:solution}, we simulate a group of 10 users and each user has a HVAC unit and local renewable generation.

\subsubsection{Algorithm Convergence}
\begin{figure}[!ht]
    \centering
    \includegraphics[width=8.5cm]{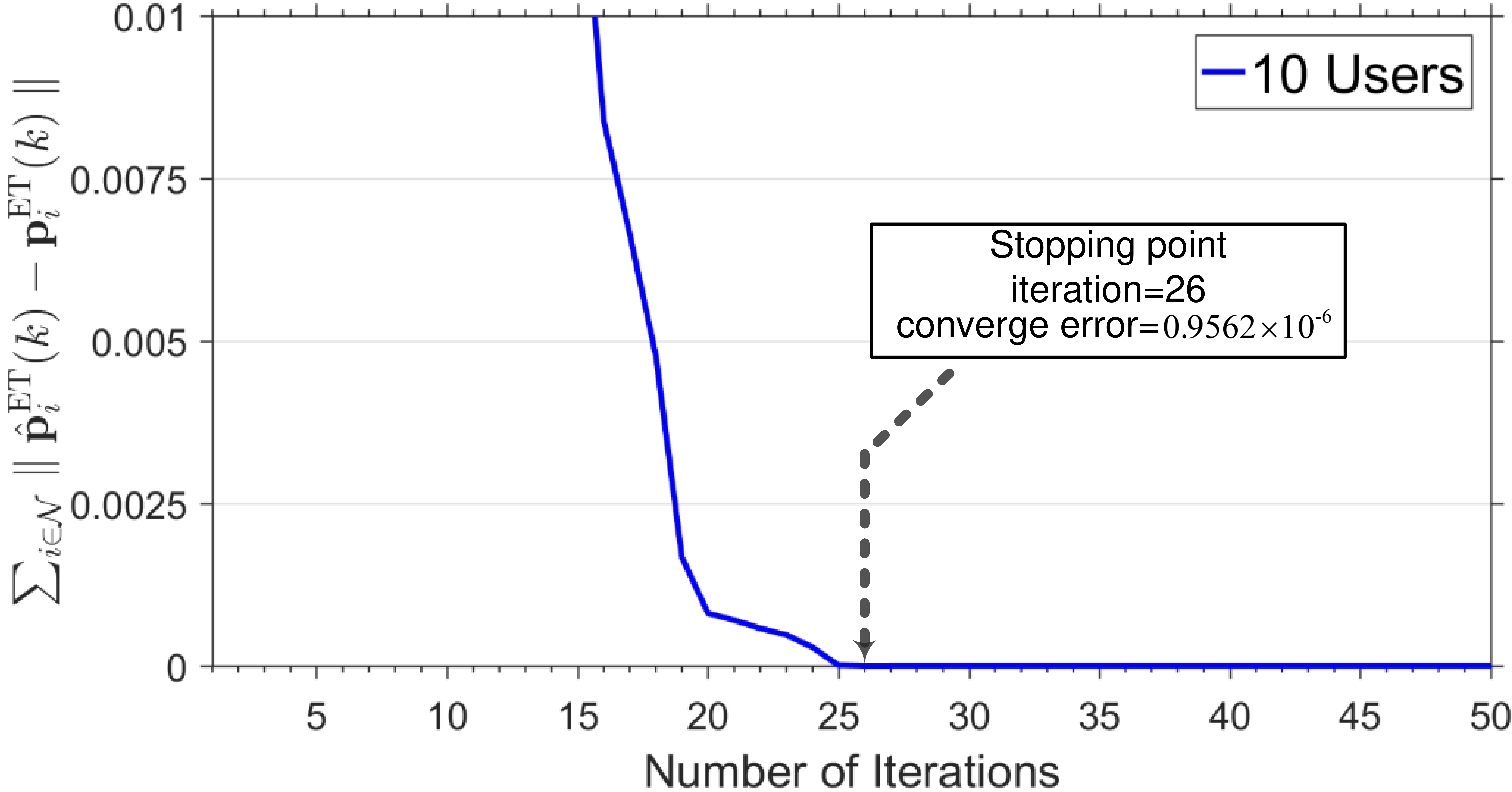}
    \caption{The convergence speed of the distributed optimization algorithm for Problem \textbf{CEMP}. We simulate $10$ users whose power usage and smart house parameters are collected from real-world data. The convergence error of the algorithm is defined in Algorithm~1, and the stop threshold is $1 \times 10^{-6}$.}
    \label{f:converge}
\end{figure}

We firstly evaluate the performance of the distributed optimization algorithm in Section~\ref{sec:solution} with 10 users. The convergence error of the distributed trading algorithms is defined as the sum of the absolute deviation between the actual trading decisions (i.e., $\bm{p}_i^{\mathrm{ET}}$) and auxiliary decisions (i.e., $\bm{\hat{p}}_i^{\mathrm{ET}}$) as in Algorithm~\ref{alg1}. We set the threshold for the convergence error to be $1 \times 10^{-6}$. We plot the error with respect to the number of iterations in Fig.~\ref{f:converge}, and we see that Algorithm~\ref{alg1} converges at iteration 26. The result show that the proposed distributed trading algorithm converges fast in practice with real-world data, which is of practical importance.

\subsubsection{Power Scheduling in Energy Trading Scenario} 
Secondly, we show the optimized hourly day-ahead energy scheduling of 10 users over one week in Fig.~\ref{f:usage} in the cooperative scenario. The decisions of users include, as shown in Fig.~\ref{f:usage}, (a) renewable energy supply, (b) the power purchased from the grid, and (c) the HVAC energy consumption of users.
We see in Fig.~\ref{f:usage} (a) that users try to utilize their renewable energy generations despite the different capacities and generations of solar energy and wind energy. In Fig.~\ref{f:usage} (b), the power purchased from the grid shows a complementary pattern with the renewable power supply, as the users try to reduce the dependence on the grid and thus reduce their costs. As depicted in Fig.~\ref{f:usage} (c), the scheduled HVAC load varies with the temperature change to keep the indoor temperature within the comfort zones.
\begin{figure*}[!b]
    \centering
    \includegraphics[width=12cm]{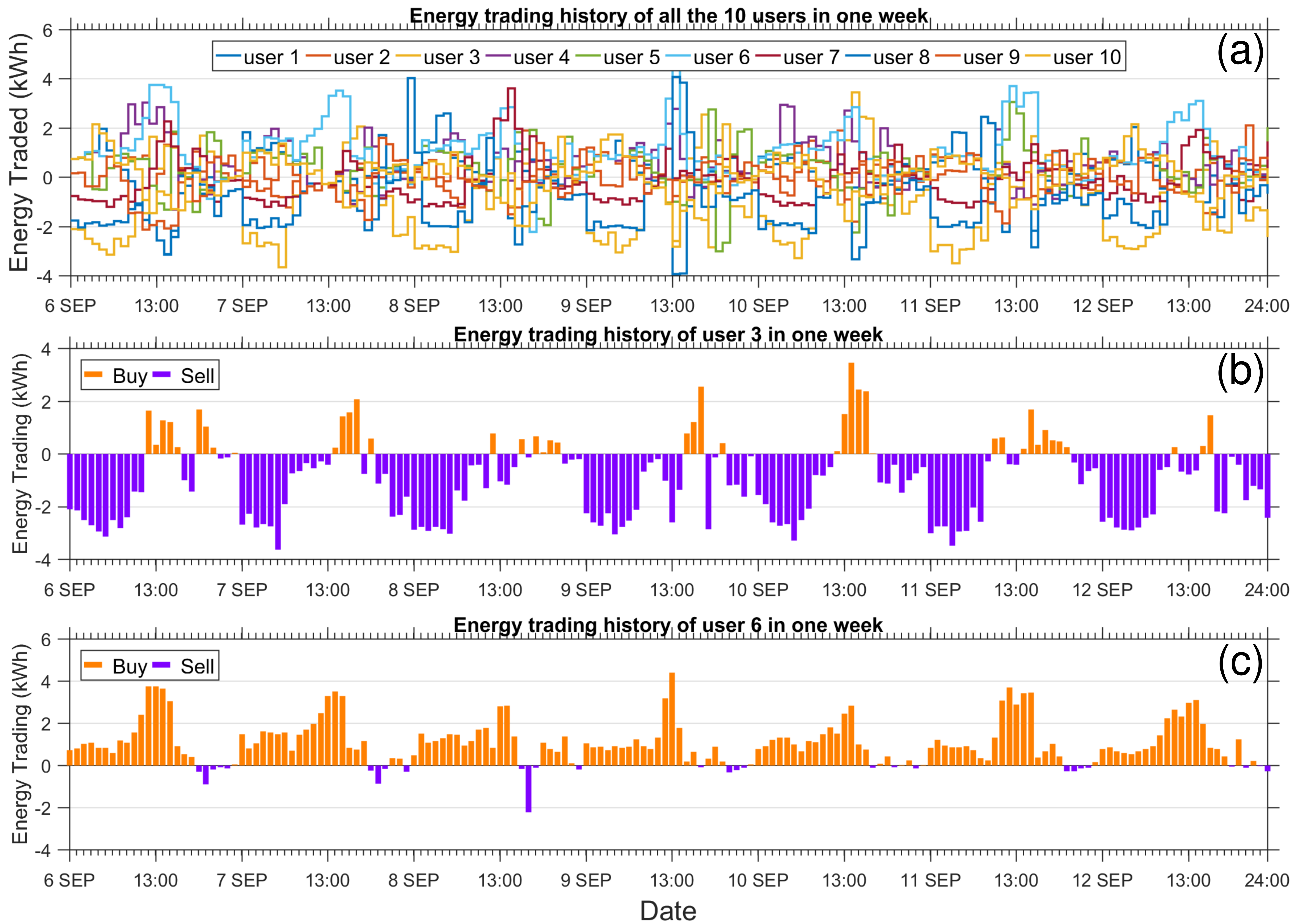}
    \caption{The optimal energy trading of $10$ users over one week: (a) the total amount of energy traded per hour for all users, (b) energy trading of user \#3, and (c) energy trading of user \#6. Note that the positive value denotes energy purchase and the negative value denotes energy selling. }
    \label{f:trading}
\end{figure*}

\begin{figure*}[!b]
    \centering
    \includegraphics[width=12cm]{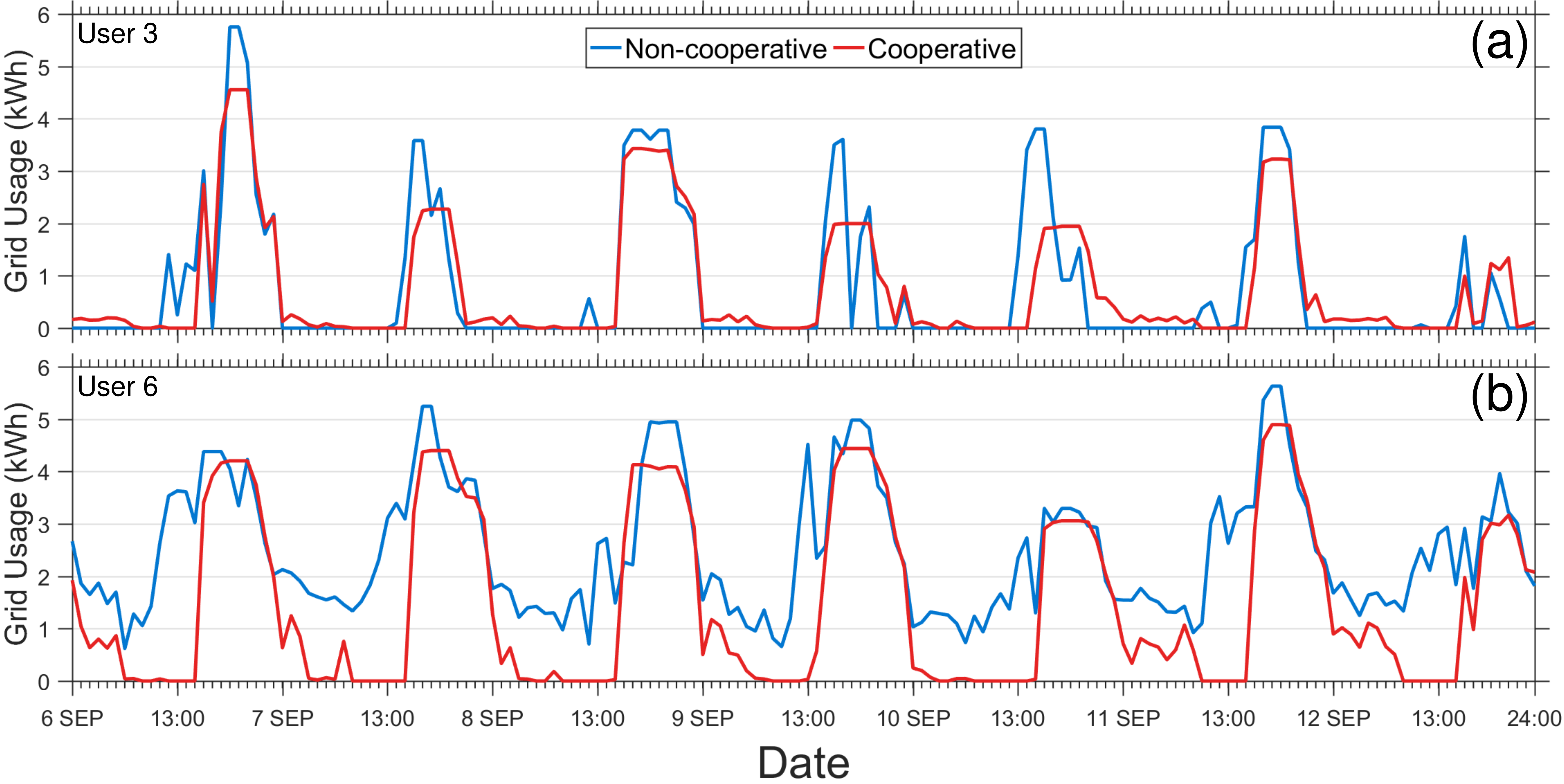}
    \caption{Comparison of the grid-purchased power between the non-cooperative and cooperative energy management for (a) user \#3 and (b) user \#6 over one-week time. We compare the non-cooperative scenario (blue) and the cooperative scenario (red) in each subplot.}
    \label{f:compare}
\end{figure*}

\subsubsection{Energy Trading Performance} 
We plot the hourly energy trading decisions of all the 10 users over one week in Fig.~\ref{f:trading} (a). We see active energy trading among users throughout the whole week, in which positive values denote energy purchase, and negative values denote energy selling. Furthermore, we take user \#3 and user \#6 as an example to show the different trading patterns of different types of users in Fig.~\ref{f:trading} (b) and (c). User \#3 is energy self-sustained and thus sells a lot of energy to other users including user \#6. 
By contrast, user \#6 is in short of local energy in the daytime and thus often purchases energy from other users.
It is demonstrated that our cooperative energy management platform provides an effective platform for users to interact and trade energy for mutual benefits. We compare the gird-purchased power of user \#3 and user \#6 as an example to show the effects of energy trading. As shown in  Fig.~\ref{f:compare}, both the users' grid power consumption, particularly the peak power consumption, is effectively reduced in the cooperative scenario compared with that of the non-cooperative scenario.
\begin{figure}[ht]
    \centering
    \includegraphics[width=8.6cm]{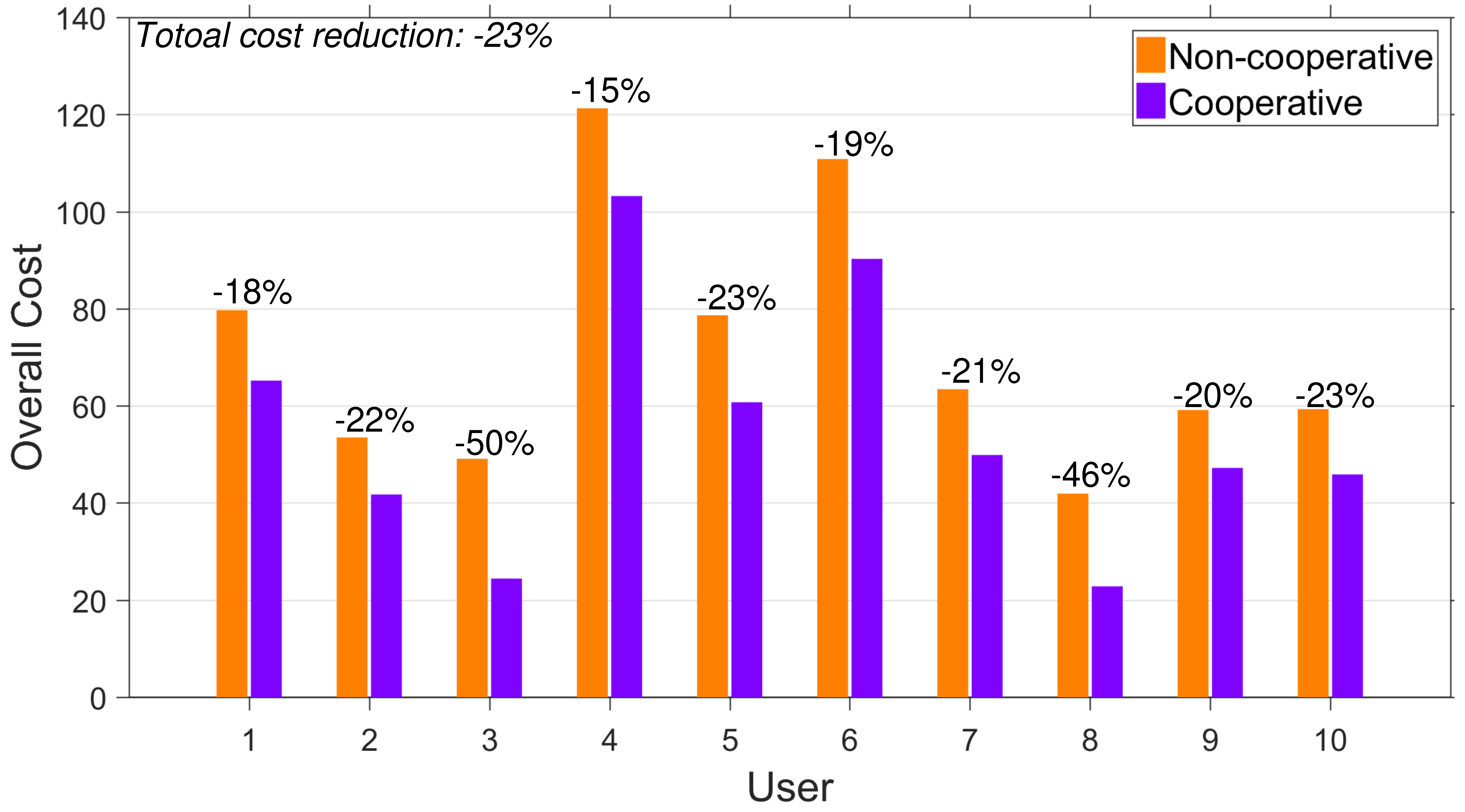}
    \caption{The costs of the ten users in the one-week simulation time. For each user, we compare its costs of non-cooperative energy management scenario (orange bar) and cooperative energy management scenario (purple bar). The numbers above the bars show the cost reduction rate for all the users.}
    \label{f:overall_cost}
\end{figure}

\subsubsection{Energy Cost Reduction}
To evaluate the benefits for users to join the cooperative energy management platform, we compare each user's cost with and without energy trading. As depicted in Fig.~\ref{f:overall_cost}, the orange bars indicate the costs of all users if they manage their power usage via Problem \textbf{EMP} in the non-cooperative scenario. The purple bars indicate the users' costs in the cooperative scenario according to Problem \textbf{CEMP}. Comparing the costs in both scenarios, we see that all the users reduce their costs directly with energy trading. We annotate the percentage of cost reduction between the non-cooperative and cooperative scenarios above the bars. It is shown that all users' costs are reduced, and the reduction rates range from 15\% to 50\%, and the total cost of the system including all the users is reduced by 23\%.

\section{CONCLUSION AND FUTURE WORK}\label{sec:conclusion}
This paper presented a cooperative energy management platform that enables residential users to trade energy with each other to improve the efficiency of their HVAC energy management. We designed a distributed energy trading algorithm that well preserves users' privacy and encourages users to trade energy to reduce energy costs. We evaluated the distributed energy trading algorithm by extensive simulations using realistic data trace. Simulation results showed that the trading algorithm reduced the costs for all the users ranging from 15\% up to 50\%, and reduced the system cost (defined as the overall cost of all the users) by 23\%. During the research of this work, we found that the proposed distributed trading algorithm well matches the structure of blockchain. In our future, we will explore to integrate blockchain technology with cooperative energy management.

\bibliographystyle{IEEEtran}
\bibliography{ref.bib}

\end{document}